\documentclass[runningheads]{llncs}
\usepackage[nottoc]{tocbibind}
\RequirePackage{chapterbib}
\usepackage{stackrel,amssymb}
\usepackage{amsmath}
\usepackage{xcolor}
\usepackage{adjustbox}

\DeclareUnicodeCharacter{2212}{\textendash}

\usepackage{listings}
\usepackage{xcolor}

\usepackage[left=1.25in,right=1.25in,top=1in,bottom=1in]{geometry}
\usepackage{url}
\usepackage{hyperref}
\usepackage{array}
\usepackage{appendix}

\newcolumntype{M}[1]{>{\centering\arraybackslash}p{#1}}

\usepackage{graphicx}
\begin{document}
\title{Using a Fast Adaptive Function Approximator to calculate Protein-Filament Binding Kinetics}
%
%
\author{Zihan Zhang\inst{1}, 
Adam Lamson\inst{2} \and Robert Blackwell\inst{3} }
\authorrunning{Z. Zhang et al.}
%
\institute{Department of Applied Mathematics, University of Washington, \\ Seattle WA 98195, USA \\ \email{zihan16@uw.edu} \and
Center for Computational Biology, Flatiron Institute, \\ New York NY 10010, USA \\ \email{}
\and
Scientific Computing Core, Flatiron Institute, \\ New York NY 10010, USA \\
}
\maketitle              
\begin{abstract}
The cytoskeleton, consisting of biopolymer filaments, molecular motors, and passive crosslinking proteins, provides the internal structure of cells that facilitate movement, growth, and cell division. Understanding the microscopic motor-filament kinetics and dynamics is essential for comprehending macroscopic behaviors of reconstituted cytoskeletal assemblies, such as self-organized flow and active stress. In this study, we employ an adaptive fast Chebyshev approximator based on tree search and parallel computing to accurately recover the equilibrium distribution of crosslinking proteins. Therefore, it satisfies detailed balance in binding through kinetic Monte Carlo sampling while maintaining cost-effectiveness. Additionally, we offer expandable features, including segregating the simulation process via pre-building and allowing the free-loading of different closed-form formulations of the motor's potential energy. Overall, this research contributes to computational advancement in function approximation and has the potential to better describe the evolution of cytoskeletal active matter. 

\end{abstract}
\newpage
\section{Introduction}
Living systems are built hierarchically, with large functional structures being composed of smaller assemblies. One classical example is the cytoskeleton, which is composed of polymer filaments, progressively walking or diffusing molecular motors, and other accessory proteins \cite{howard}. It remains unclear how diverse cytoskeleton structures, including the mitotic spindle, cortex, and flagella, dynamically assemble and generate forces in different magnitudes \cite{barnhart, bornens, polland}. 

Within the cytoskeletal framework, the processes of force generation and reorganization depend on crosslinking motors, as they align and slide filament pairs \cite{lamson2021}. For example, during mitosis and chromosome segregation, kinesin and dynein motors crosslink and rearrange microtubules to rapidly assemble the mitotic spindle \cite{redemann, she2017, McIntosh}. Similarly, myosin motors reorganize actin networks to significantly change a cell's morphology during cell crawling or cytokinesis \cite{gupton2006, laevsky2003, fournier2010}. Gaining a comprehensive understanding of the dynamic and mechanical attributes of cytoskeletal filament networks is imperative. This knowledge equips researchers to anticipate the causal connections between molecular perturbations and cellular responses, fostering the development of adaptive and versatile materials \cite{li2008, needleman2017}.  

The \texttt{aLENS} (a Living ENsemble Simulator) framework was developed for efficiently and accurately simulating a large-scale cytoskeletal system of $N$ grid bodies interconnected by dynamic springs \cite{aLNEs}. In essence, protein motors undergo processes of binding, unbinding, or crosslinking with filaments. To maintain kinetic detailed balance both globally (for the entire population) and locally (for each object) in the passive limit, aLENS simulates motors as navigating a precisely defined free energy landscape and prevents the occurrence of artificial energy flux \cite{lamson2021}. During each time step, the algorithm sequentially addresses three challenges: \textbf{a)} motor diffusion and stepping, \textbf{b)} computation of binding and unbinding events while preserving realistic macroscopic statistics, and \textbf{c)} updating filament positions, surmounting stiffness constraints, and ensuring steric exclusion \cite{aLNEs}. 


In our current study, we introduce an adaptive fast function approximator named \texttt{Baobzi} based on tree-search methods to enhance the resolution of \textbf{b)} tasks. This approach exhibits broad applicability across diverse approximation challenges and proves particularly advantageous within the context of this low-dimensional inverse probability integral transform scenario. With the help of this computational tool, we attain a more nuanced characterization of motor binding probability with increased efficiency. This advancement holds substantial promise for generating macroscopic simulations that closely align with real-world phenomena. Additionally, we offer more flexible user interfaces, delineating the simulation workflow through pre-built partitions and enabling the integration of various adaptable formulations of motor potential energy.

\section{Kinetic Monte-Carlo (KMC) method}
We utilize the Kinetic Monte-Carlo method to simulate interactions between filaments and crosslinking proteins as passive processes  \cite{gao2015a, blackwell2016, lamson2019, chris2020, lamson2021}. Specifically, we could formulate the motor's binding condition with $4$-states: neither head attached to a filament ($U$), bound with one head attached to a filament ($S_A, S_B$), and crosslinking two filaments ($D$). Transition rates $R(t)$ ensure the correct limiting distribution by enforcing detailed balance, and the transition probabilities are modeled as inhomogeneous Poisson processes: 
\begin{equation}
    P(\Delta t) = 1 - \exp\left(-\int_0^{\Delta t} R(t)dt\right) = 1- \exp(-R(0)\Delta t + O(\Delta t^2)).
\end{equation}
The transition $(S_A, S_B) \leftrightarrows D$ maintains macroscopic thermodynamic statistics, ensuring accurate equilibrium bound-unbound concentrations and distributions, while also considering the energy associated with tether deformation. When one head is bound to a filament $i$ at position $s_i$, the unbinding rate and total binding rate (calculated as an integral of Boltzmann factors) are:
\begin{equation}\label{bindingrate}
\begin{aligned}
R_{\text{on},D}(s_i,t) &= \dfrac{{\epsilon K_e k_{o,D}}}{V_{\text{bind}}}\sum_j \int_{L_j} e^{-\beta U_{i,j}(s_i,s_j)}ds_j, \\
R_{\text{off},D}(s_i,s_j,t) &= k_{o,D}. 
\end{aligned}
\end{equation}
where $\beta=(k_B T)^{-1}$ is the inverse temperature, $k_{o, D}$ is the force-independent off-rate constant, $K_e$ is the effective binding association constant, and $U_{i,j}$ is the free energy contribution from the tether \cite{aLNEs}: 
\begin{equation}\label{energy}
    U_{i,j} = \dfrac{\kappa_{xl}}{2}(\ell(s_i,s_j)-\ell_0 - D_{\text{fil}})^2. 
\end{equation}
where $D_{\text{fil}}$ is the filament's diameter. $V_{\text{bind}}$ is the exploring volume centered at the unbounded motor: 
\begin{equation}
    V_{\text{bind}} = \int e^{-\beta U_{i,j}}dr^3 = 4\pi \int_0^{R_{\text{cut},C}}e^{-\beta U_{i,j}}r^2dr.
\end{equation}
For approximation, the integrand becomes small beyond the cutoff radius $R_{\text{cut}, C}$. $L_j$ is the segment of filament $j$ embedded in $V_{\text{bind}}$\footnote{We may extend $L_j$ to set $\{ L_k \}_{k \in \mathcal{A}}$ and linearly sum each term if there are multiple filaments included in $V_{\text{bind}}$ of $s_i$.}. Our model's framework could also incorporate force dependency into the binding rate. This integration facilitates catch-bond-like behavior, enabling proteins to maintain crosslinks for extended periods under tension while promoting quicker release when compressed: 
\begin{equation}\label{forcedependence}
    f_F(s_i,s_j) = \kappa_{\text{xl}}\left(\dfrac{\lambda}{2}(\ell(s_i,s_j)-\ell_0)^2 + x_c(\ell(s_i,s_j)-\ell_0) \right)
\end{equation}
where $\lambda$ and $x_c$ represent the energy factor and characteristic length, respectively, defining the dynamics of energy- and force-dependent binding/unbinding. For values of $x_c <0$, catch-bond-like behavior is observed, whereas $x_c>0$ results in slip bond behavior \cite{walcott, chris2020}. By incorporating Equation (\ref{forcedependence}) into the integrand of Equation (\ref{bindingrate}), we could modify the formulation of the (un)binding rates. This alteration does not affect the final stored energy in either bound state but influences the frequency at which the motors alternate between single-head binding and crosslinking \cite{aLNEs}. For the scope of this paper, we focus exclusively on the potential-based formulation, highlighting the model's adaptability to more complex formations with more free parameters. 

Direct numerical integration in Equation (\ref{bindingrate}) at each time step is time-consuming, so the \texttt{Lookup Table} approach is proposed \cite{lamson2021}. We achieve dimensionality reduction of the cumulative density function (CDF) by incorporating the lab position of each bound motor head and an infinite carrier line defined by the position and orientation of the unbound filament. Therefore, binding is determined by the minimum distance $r_{\perp}$ between the bound motor head position and the filament ends $[s_-, s_+]$ on the carrier line: 
\begin{equation}\label{actual_CDF}
    \text{CDF}(r_{\perp},s) = \text{sgn}(s) \int_0^s e^{-\beta U(r_{\perp},s')} ds' := c. 
\end{equation}
Therefore, the total crosslinking rate on $L_j$ is determined as the difference between two corresponding CDFs using its starting and ending positions. We calculate the benchmark values of Equation (\ref{actual_CDF}) through Gauss-Konrad integration and create a lookup table using uniformly discretized 2D meshes across the domain:
\begin{equation}\label{domain}
    s, r_{\perp} \in \left[ 0, \sqrt{-\dfrac{2\ln(\delta)}{\beta k_{\text{cl}}}} 
 + h_{\text{cl}} \right ]
 \end{equation}
where $\delta$ is the accuracy limit, $h_{\text{cl}}$ is the modified tether length, and $k_{\text{cl}}$ is the effective tether spring constant \cite{lamson2021}. We use the 2D linear interpolation for inputs $r_{\perp}$ and $s$ in normal lookup, together with binary search with complexity $O(\log_2(\delta s_{\max}))$ for inputs $r_{\perp}$ and $c$ through inverse transformation sampling \cite{lamson2021, aLNEs}. Hence, lookup tables, tailored to motor and filament types, are prebuilt just once. Subsequently, the simulation process only requires function evaluations. Smaller prespecified grid spacings ($\Delta s, \Delta r$) would decrease approximation errors at the expense of increased memory usage and computation time. Additionally, the algorithm performs badly when the CDF exhibits a low slope (nearly horizontal) at large $s$ values and when the corresponding probability density function (PDF) resembles the Dirac delta function (Figure \ref{pos_search_result}). Repetitive errors in small-scale interactions can accumulate, leading to significant divergences at the macroscopic level and deteriorating the simulations. Hence, a more adaptive and efficient framework is essential.

\section{Baobzi Algorithm}
\begin{figure}[!htb]
    \centering
    \includegraphics[width=1\textwidth]{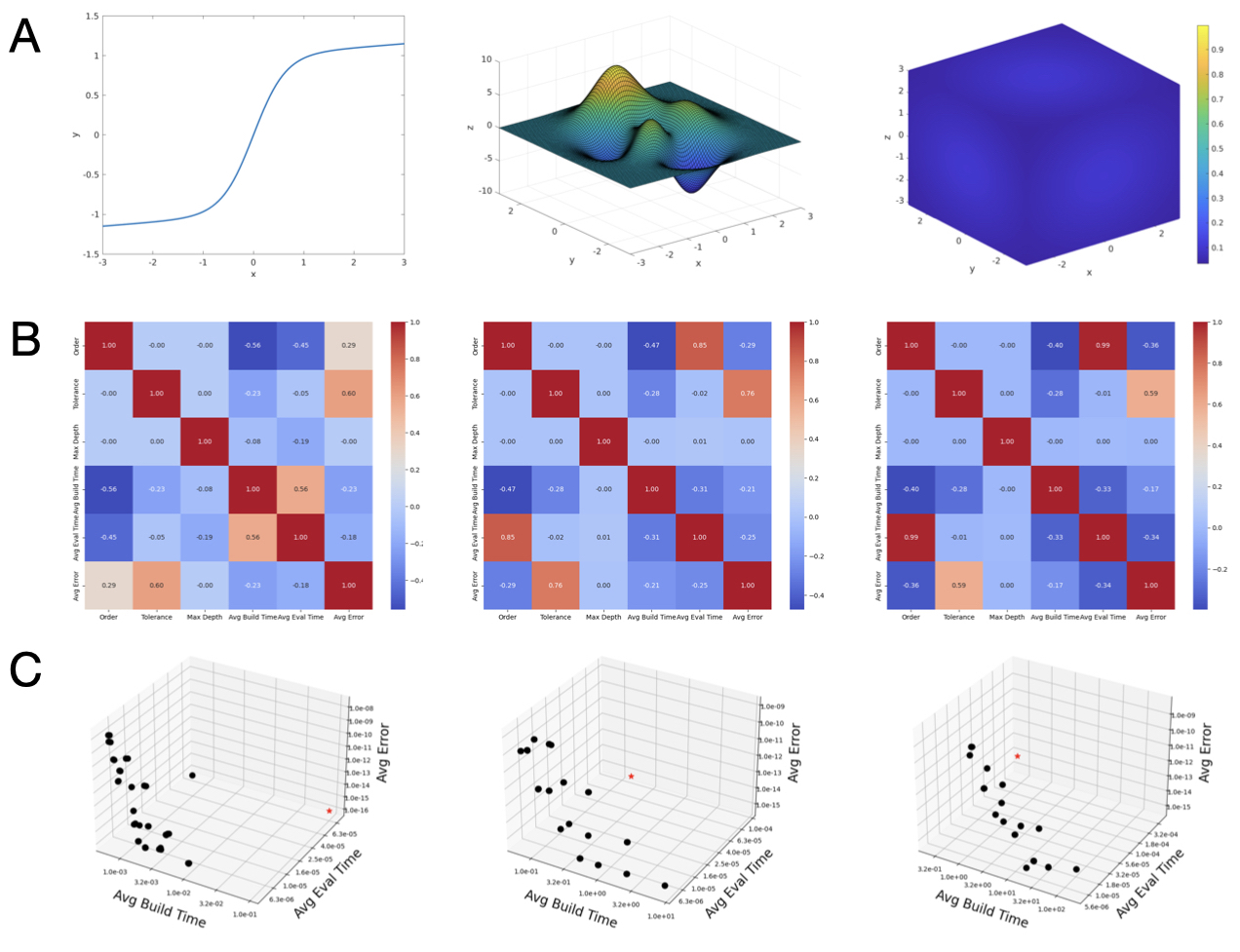}
    \caption{Comparative analysis of \texttt{Baobzi} (MATLAB version) and \texttt{Chebfun} algorithms across multiple dimensions. Examples are adapted from \texttt{Chebfun} documentation \cite{Driscoll2014}. (A) Function illustrations for 1D, 2D, and 3D inputs within the domain $[-3,3]^n$. The 1D case presents $f(x) = \tanh(\pi x/2) + x/20$, the 2D case uses the \texttt{peaks} function, and the 3D case employs $f(x,y,z)=1/(1+x^2+y^2+z^2)$, with spatial color coding. (B) Correlation of hyperparameters (\textit{order}, \textit{tol}, \textit{max\_depth}) with outcomes (approximation error, build time, evaluation time). Notably, stricter tolerance improves performance but increases build time. Higher-order polynomials enhance accuracy but may cause Runge's Phenomenon in simpler cases. The $max\_depth$ parameter has minimal impact. (C) Performance comparison of \texttt{Chebfun} (red point) and \texttt{Baobzi} (black points) algorithms. \texttt{Baobzi}, with parameter optimization, potentially surpasses \texttt{Chebfun} in efficiency, as indicated in the lower-left quadrant. Better performance is expected if using other low-level and compiled languages. }
    \label{baobzi_cheb}
\end{figure}

We design the \texttt{Baobzi} algorithm, employing tree search to convert CPU-intensive function calculations into computationally cheaper alternatives, albeit with memory costs. Internally, \texttt{Baobzi} represents the function by a grid of binary/quad/oct/N trees, where the leaves represent the function in some small sub-box of the function's domain with Chebyshev polynomials. The Clenshaw algorithm is used for function summations. For point evaluation, it searches the tree for the box containing this point and evaluates using this approximant. An exemplary application could be: building a complicated or high computational cost function in a higher-level language, constructing a \texttt{Baobzi} function, using it as a drop-in replacement to your function (reaping speed benefits), then taking a prior function, exporting it, and using it in a high-performance language (for production or prototype). With its straightforward \texttt{C} interface and compatibility with languages such as \texttt{C++}, \texttt{Fortran}, \texttt{Python}, \texttt{Julia}, and \texttt{MATLAB}, \texttt{Baobzi} demonstrates remarkable versatility and adaptability. It also does not require other library dependencies and has CPU dispatch with the automatic setting of optimized codepath without any user intervention. Key hyperparameters of \texttt{Baobzi} creation include domain parameters (\textit{center}, \textit{half\_length}) and input structure parameters (\textit{dim}: [number of independent variables]; \textit{order}: [polynomial order]; \textit{tol}: [maximum desired relative error]; \textit{max\_depth}: [maximum depth allowed for tree before interpolation considered failed]). The proposed algorithm shares similarities with \texttt{Chebfun} in MATLAB, renowned for its utilization of piecewise polynomial interpolants and Chebyshev polynomials. However, it offers a notable enhancement in computational efficiency, which is achieved by employing polynomial fits of considerably lower order while maintaining comparable accuracy and tolerances \cite{chebfun, platte2010, Driscoll2014}. Figure \ref{baobzi_cheb} provides an in-depth analysis of the \texttt{Baobzi} algorithm through some representative examples, exploring its parameter sensitivity and correlation, and offers a detailed comparison with \texttt{Chebfun}. The (C) subplot distinctly showcases \texttt{Baobzi}'s superior performance across all metrics, including approximator build time and pointwise approximation error, post-parameters fine-tuning. Universally, the average pointwise approximation evaluation time using \texttt{Baobzi} (across various settings) is consistently faster than \texttt{Chebfun}, underscoring its potential for implementation in large-scale dynamic systems.

Despite the computational benefits offered by \texttt{Baobzi}, it is not without limitations that are typical in the realm of numerical analysis algorithms \cite{isaacson1994analysis, süli2003introduction}. The algorithm exhibits suboptimal performance in the vicinity of singularities or when dealing with pathological functions. It lacks an intrinsic mechanism for bound checking, necessitating users to ensure input sanitation and, where required, construct multiple piecewise functions. Additionally, \texttt{Baobzi} tends to consume significant memory resources, particularly when dealing with oscillatory or rapidly varying functions, which necessitates the application of appropriate transformations and fitting techniques. 

\section{Results}
Here, we implement the \texttt{Baobzi} algorithm in the KMC context and test its effectiveness under various environments and parameter settings. Specifically, $\alpha (\mu m^{-2}) \propto \beta$ describes the motor's stiffness, and $\ell (\mu m)$ denotes the motor's freelength. These parameters will be varied on magnitudes of $10^2$ and $10^1$ for ad-hoc tests, evaluating the model's robustness and adaptivity.

\subsection{Normal Domain Search}
\begin{figure}[!htb]
    \centering\includegraphics[width=1\textwidth]{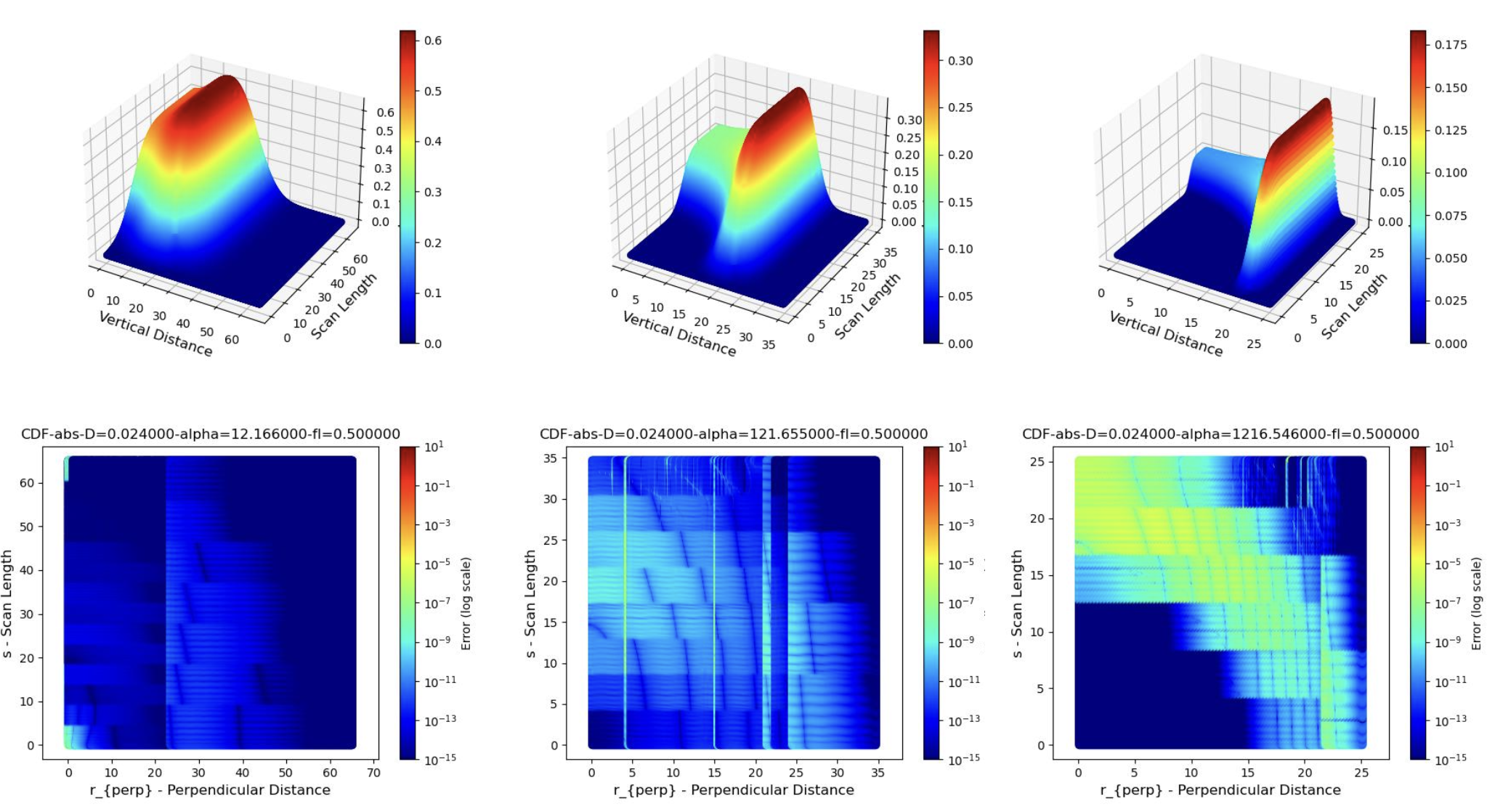}
    \caption{Comprehensive domain search on $s$ (scan length) and $r_{\perp}$ (vertical distance) within the specified lookup table domain (Equation (\ref{domain})). Approximated CDF of the multivariate normal distribution (left) and error graph (right). Examples are shown under soft (left), medium (middle), and hard (right). }
    \label{pos_search_result}
\end{figure}

\begin{figure}[!htb]
\centering
\begin{minipage}[t]{0.325\textwidth}
\centering
\includegraphics[height=4cm, keepaspectratio]{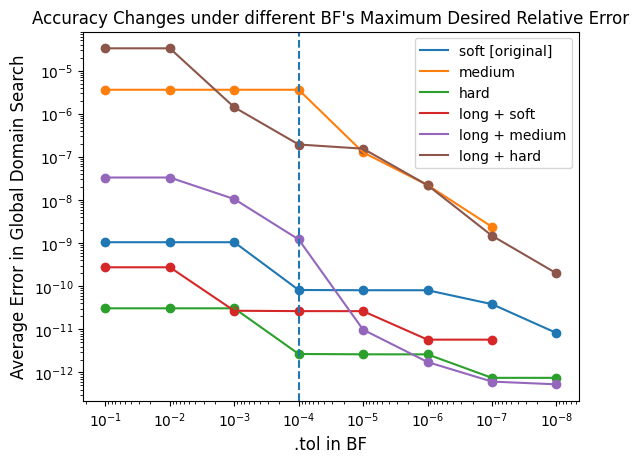}
\end{minipage}
\centering
\begin{minipage}[t]{0.325\textwidth}
\centering
\includegraphics[height=4cm, keepaspectratio]{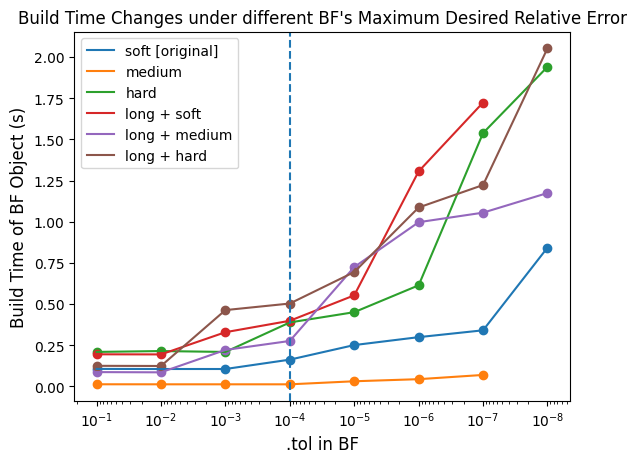}
\end{minipage}
\centering
\begin{minipage}[t]{0.325\textwidth}
\centering
\includegraphics[height=4cm, keepaspectratio]{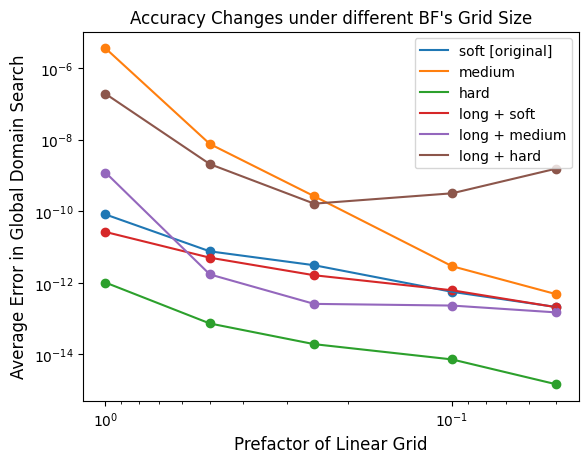}
\end{minipage}
\caption{Sensitivity analysis of BF's hyperparameters in normal lookup. Change of average pointwise error (left) and build time (middle) in global domain search under different stiffness and freelength of motor spring as functions of maximum desired relative error ($\textit{tol}$). Smaller $\textit{tol}$ takes longer build time, so balancing it with accuracy, we choose $\textit{tol} = 10^{-4}$ as a suitable default choice. Decreasing the grid size (right) could further improve BF's optimal performance in magnitudes, but also require much computing resources. }
\label{errortolchange}
\end{figure}

\begin{figure}[!htb]
    \centering
    \includegraphics[width=1.0\textwidth]{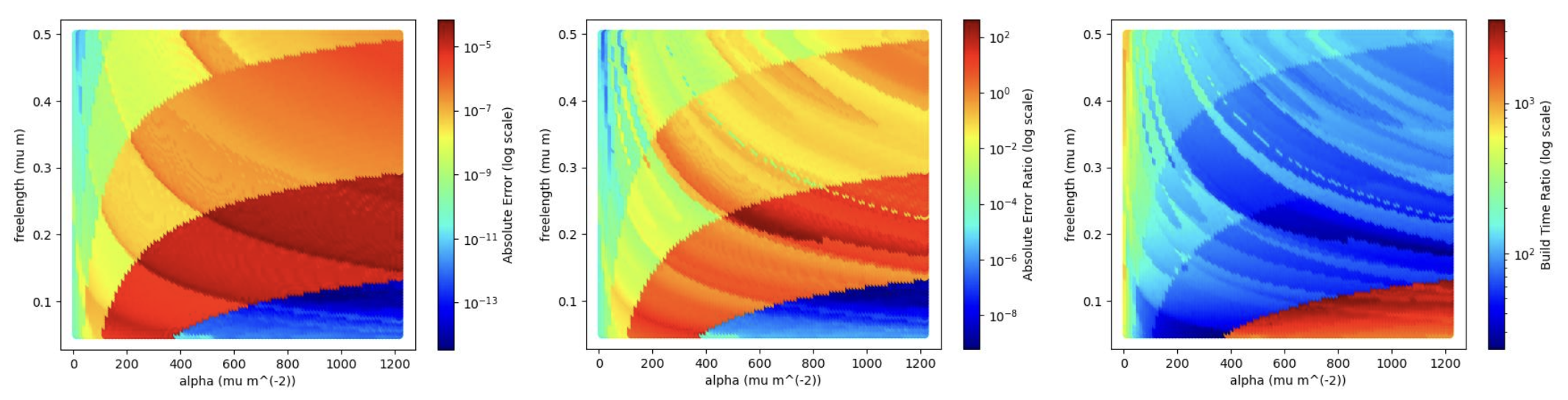}
    \caption{BF's average error (left) and comparison of relative ratio (dimensionless, presented in log scale) between BF and BF2 on accuracy (middle) and build time (right) in normal lookup by iterating through $10^4$ different combinations of $\alpha$ and $\ell$. Object parameters are identically set, including $\textit{tol}=10^{-4}$ and BF's grid coefficient $a=1$. }
    \label{bfbf2compare}
\end{figure}

The workflow includes linear discretization of the $r_{\perp}$ dimension, generating multiple \texttt{Baobzi} objects using the target function and domain parameters, which are saved into a vector (named \texttt{Baobzi Family} (BF) object). Binary search is employed for evaluating specific points. In Figure \ref{pos_search_result}, an exemplary comprehensive domain search is performed over the 2D field, and pointwise relative errors are calculated. Table \ref{baobzilookupcomparison} illustrates a comprehensive comparison, highlighting BF's superior interpolation accuracy at a reasonable and affordable cost. Furthermore, the model demonstrates consistent performance even under high motor stiffness, resulting in a sharp bump function for the PDF. In Figure \ref{errortolchange}, we compute and compare the accuracy under various maximum desired relative errors and choose $\textit{tol}=10^{-4}$ as a balanced optimal choice for normal lookup under all circumstances. Any reasonably large, even-order setting for \textit{order} will effectively fulfill the approximation, with the default choice set to 8. By experience, the approximation would be the most efficient if the half lengths along both dimensions are within the same magnitude ($10^n$ for $n \in \mathbb{Z}^-$). For better accuracy but with exponentially worse demanding space requirements and build time costs, the grid size could be smaller with some coefficients $a$ (factor of $10$) multiplications. Another intuitive approach (named BF2) is to normalize the entire square domain and create a single \texttt{Baobzi} object, which reduces build time, worsens accuracy, and maintains evaluation efficiency with the same setting of parameters. In Figure \ref{bfbf2compare}, we present a quantitative and systematic comparison between BF and BF2 across a broad range of parameters. The approximator construction only needs to be conducted once through pre-building, which diminishes the advantages of BF2 and leads us to favor using BF.

\begin{table}[!htb]
\caption{Comparison of performances between \texttt{Lookup Table} (LT) and \texttt{Baobzi Family} (BF) under different stiffness and freelength of motor spring. LT's dimension is $256 \times 256$. \textit{Long} indicate $10$ times of freelength and the magnitude difference between \textit{soft/medium/hard} on $\alpha$ is $1/10$ respectively. Generally, BF would have much more accurate measurements with longer time costs both in creating the object and point evaluations ($\sim 10^6$ different instances). BF's creation is performed once before the KMC operation, minimizing its impact on time costs. }
\centering
\begin{tabular}{|M{3.5cm}|M{2.1cm}|M{2.1cm}|M{2.1cm}|M{2.1cm}|M{2.1cm}|M{2.1cm}|}
\hline
& \text{Soft} & \text{Medium} & \text{Hard} & Long\&Soft & Long\&Medium & Long\&Hard \\
\hline
LT Global Test Acc & 3.548e-07 &  7.280e-06 & 2.749e-06 & 9.596e-07 & 5.653e-07 & 6.067e-06 \\
\hline
\textcolor{red}{BF Global Test Acc} & \textcolor{red}{8.185e-11} & \textcolor{red}{3.599e-06} & \textcolor{red}{1.020e-12} & \textcolor{red}{2.651e-11} & \textcolor{red}{1.216e-09} & \textcolor{red}{1.935e-07} \\
\hline
LT Build Time (s) & 0.022 & 0.023 &  0.022 &  0.021 & 0.028  & 0.079 \\
\hline
BF Build Time (s) & 0.170 &  0.013 & 0.406 & 0.404 & 0.279  & 0.508 \\
\hline
\textcolor{red}{Build Time Ratio} &  \textcolor{red}{7.644}  &  \textcolor{red}{0.568}  &  \textcolor{red}{17.740} &   \textcolor{red}{18.828} & \textcolor{red}{10.028}  &  \textcolor{red}{6.448} \\
\hline
LT Evaluation Time (s) & 0.014 & 0.005
 & 0.002 &0.0219 & 0.012 & 0.009 \\
\hline
BF Evaluation Time (s) &   0.131 &  0.045 & 0.057 & 0.260 & 0.107 & 0.134 \\
\hline
\textcolor{red} {Evaluation Time Ratio} &  \textcolor{red}{8.935} & \textcolor{red}{8.852} &  \textcolor{red}{26.763} & \textcolor{red}{11.875}  &  \textcolor{red}{8.696}  & \textcolor{red}{14.363} \\
\hline 
\end{tabular}
\label{baobzilookupcomparison}
\end{table}

\subsection{Reverse Lookup Search}
\begin{figure}[!htb]
\centering
\begin{minipage}[t]{0.49\textwidth}
\centering
\includegraphics[width=7cm]{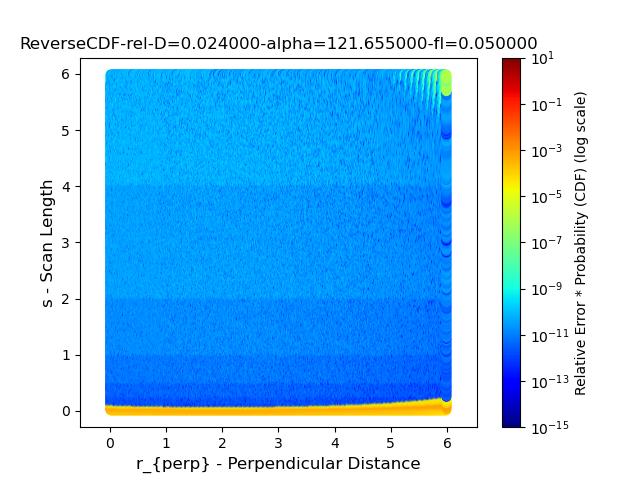}
\end{minipage}
\centering
\begin{minipage}[t]{0.49\textwidth}
\centering
\includegraphics[width=7cm]{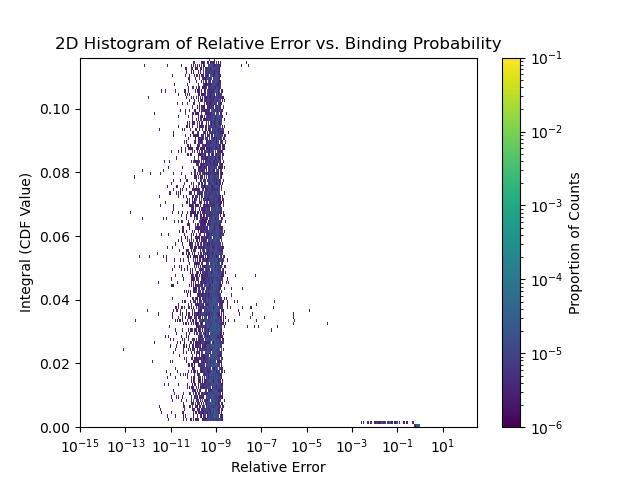}
\end{minipage}
\caption{Relative error scaled with binding probability through element-wise dot products (left) and 2D histogram between relative error and binding probability (right) when coefficient $a=0.1$. Most large errors are distributed when the integral value is small. Except the motor is exceptionally hard, most errors are condensed in $10^{-9}$ to $10^{-10}$ level. }
\label{histogram_plot}
\end{figure}

\begin{figure}[!htb]
\centering
\begin{minipage}[t]{0.49\textwidth}
\centering
\includegraphics[height=4cm, keepaspectratio]{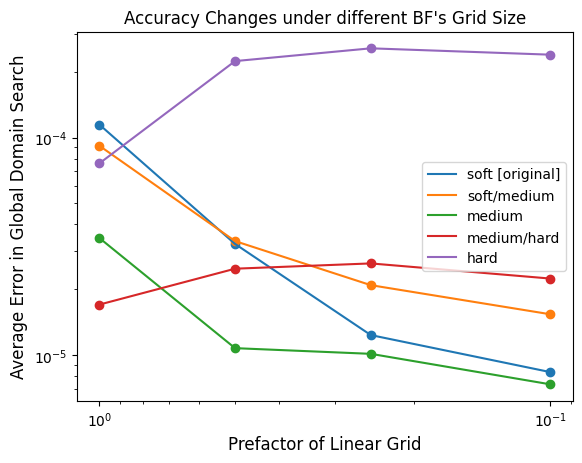}
\end{minipage}
\centering
\begin{minipage}[t]{0.49\textwidth}
\centering
\includegraphics[height=4cm, keepaspectratio]{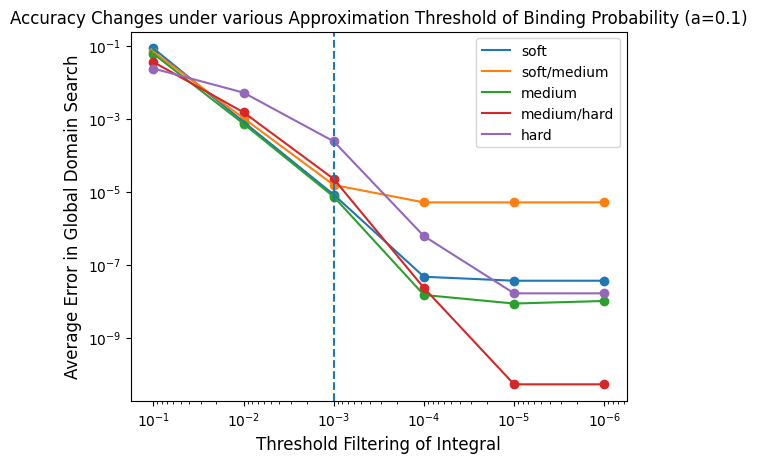}
\end{minipage}
\caption{Sensitivity analysis of BF's hyperparameters in reverse lookup under different coefficients $a$ of the linear grid (left) and under various binding probability filtering thresholds when $a=0.1$ (right). }
\label{threshold}
\end{figure}

\begin{figure}[!htb]
    \centering
    \includegraphics[width=7cm]{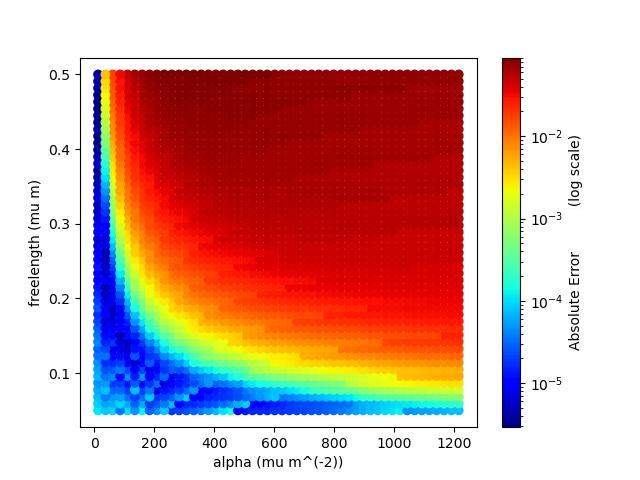}
    \caption{BF's average error in reverse lookup by iterating through $10^4$ different combinations of $\alpha$ and $\ell$. }
    \label{reverse_data}
\end{figure}

Similarly, we conduct a reverse lookup search over the whole domain where $s$ is subject to be calculated where $r_{\perp}$ and $c$ are given. We first precalculate $c$'s range in each \texttt{Baobzi} object on the linearly discretized grid, then match its grid size along $r_{\perp}$. Since CDF is a non-decreasing function and may only be compactly supported in a small interval, to promise good speed, we set up a threshold $c_{\text{threshold}}=10^{-3}$ to identify the integral to $0$ and approximate the constant function easily. To find the root without derivatives, we use the bisection method in the Boost \texttt{C++} library, while other approaches including Bracket and TOMS748 are also effective \cite{Alefeld1995Algorithm7E}. Creating each BF object is mutually independent, therefore we utilize parallel computing like OpenMP to accelerate the process. Table \ref{reverselookupcomparison} describes the details of BF's performance under the loose standard in different settings. Figure \ref{histogram_plot} shows the statistical distributions of errors, indicating that most failures in approximation occurred in scenarios with inherently low total binding probabilities (small $s$), which are unlikely to occur in reality and thus weaken the significance of the problem. In Figure \ref{threshold}, we examine the model's sensitivity on $a$ and $c_{\text{threshold}}$. Compared to the outcomes in Table \ref{reverselookupcomparison}, the model's approximation accuracy can be significantly enhanced at the expense of increased time and space costs. Given the potentially lengthy build time for reverse lookup, we offer an API to save the BF object as individual external files, allowing reconstruction. This separation of the pipeline into build-up and evaluation stages offers more flexible time management. Table \ref{reconstructdata} presents a detailed performance record of the reconstruction process. Another development is simply using the effective PDF region and discarding the low-probability binding sites to reduce the search area. The region is diagonally symmetric so an annulus or sector could be formed and analyzed through conformal mapping \cite{conformal, conformal2, reverseannulus}. In Figure \ref{reverse_data}, we thoroughly test BF's performance in reverse lookup and find BF's advancement in accuracy particularly when $\alpha$ and $\ell$ are small.

\begin{table}[!htb]
\caption{Comparison of performances between \texttt{Lookup Table} (LT) and \texttt{Baobzi Family} (BF) in reverse lookup test. $\textit{tol}=10^{-1}$ and grid size is $0.01$. Both criteria are loosely constrained to ensure test efficiency across all scenarios. The proportion of $\alpha$ value between each column is $0.1:0.5:1:5:10$.  All parameters are chosen to optimize build time, not accuracy, in this table. The accuracy results are very similar between the two methods under the current test conditions.}
\centering
\begin{tabular}{|M{4cm}|M{2.2cm}|M{2.2cm}|M{2.2cm}|M{2.2cm}|M{2.2cm}|}
\hline
& \text{Soft} & Soft/Medium & Medium & Medium/Hard & Hard \\
\hline
\textcolor{red} {LT Global Test Accuracy} & 6.771e-05 &   2.901e-05 &  1.957e-05 & 5.701e-06 &  2.470e-06 \\
\hline
\textcolor{red} {BF Global Test Accuracy} &  1.147e-04 &   9.168e-05 &  3.451e-05 &  1.703e-05 & 7.599e-05 \\
\hline 
\textcolor{red} {BF Relative Error} & 0.164\% &  0.268\% & 0.288\% & 0.569\% & 3.481\% \\
\hline 
BF Build Time (s) & 24.336 & 9.196 & 6.109 & 3.204 & 2.271 \\
\hline 
BF Required Space (MB) & 195.410 & 92.759 & 68.968 & 33.767 &  24.675 \\
\hline 
\end{tabular}
\label{reverselookupcomparison}
\end{table}

\begin{table}[!htb]
\caption{Required time and space to reconstruct BF object under different settings. }
\centering
\begin{tabular}{|M{4cm}|M{2.2cm}|M{2.2cm}|M{2.2cm}|M{2.2cm}|M{2.2cm}|}
\hline
& \text{Soft} & Soft/Medium & Medium & Medium/Hard & Hard \\
\hline
Reconstruction Time (s) & 4.299 & 2.130 & 0.891 & 0.591  & 0.302  \\
\hline
Required Space (MB) & 171 & 81  & 60 &29  & 21  \\
\hline 
\end{tabular}
\label{reconstructdata}
\end{table}

\section{Conclusion}
In this research, we introduce \texttt{Baobzi}, an adaptive approximation algorithm that utilizes Chebyshev polynomials and tree structures to approximate functions with notable efficiency in terms of computation speed and precision. Leveraging parallel computing, \texttt{Baobzi} is applied to simulate motor binding rates on filaments, resulting in enhanced accuracy and cost-efficiency for both search directions. The module is multi-language compatible, providing greater flexibility in development and application. Although this work has been tested only with small-scale toy tasks, it includes a \texttt{Lookup Table} that sets benchmarks for Kinetic Monte Carlo (KMC) tests and offers potential for integration into more comprehensive large-scale cell-level simulators \cite{aLNEs}. \texttt{Baobzi} offers additional functionalities, such as process segregation through pre-building and the capability to integrate different closed-form formulations based on user inputs, like the potential-based $U_{i,j}(s_i,s_j)$ could be expanded to account for force and angular dependencies \cite{evans, Dudko2006IntrinsicRA, walcott, guo2019, aLNEs}. However, the algorithm's current limitation is its inability to scale beyond three-factor dependencies, constraining the dimensionality of input parameters. Multiple piecewise or intermediate approximators may be helpful with the potential danger of overlapped errors. Also, the algorithm faces exponential increases in time and memory requirements for higher approximation accuracies, particularly in scenarios involving Gaussian distributions with narrow bandwidths. Therefore, exploring the potential integration of rejection sampling or other Markov Chain Monte Carlo (MCMC) methods to calculate integrals could be beneficial \cite{casella2004generalized, rejsample}. 

\bigskip
\textbf{\ackname} 
This study received support from the Biophysical Modeling Group at the Center for Computational Biology, Flatiron Institute, Simons Foundation. The authors are grateful for insightful discussions with Bryce Palmer, Dr. Christopher Edelmaier, and Dr. Michael J. Shelley. This work is accepted in the poster track in Research in Computational Molecular Biology (RECOMB 2024\footnote{\url{https://recomb.org/recomb2024/}}) held in Boston, Massachusetts.

\textbf{Code Availability.} This work's source code can be found on GitHub\footnote{\url{https://github.com/StevenZhang0116/KMC/tree/master}}. The \texttt{Baobzi} toolkit is currently available online and is being actively maintained\footnote{\url{https://github.com/flatironinstitute/baobzi}}. 

\newpage
\appendix
\section{Demo Code}

Here we provide a simple demo code in \texttt{C++} for constructing and reconstructing the BF object for both normal and reverse lookup. The function APIs mimic those from the previously created lookup table \cite{lamson2021}\footnote{\url{https://github.com/lamsoa729/KMC}}.

\subsection{Reverse Lookup}

\begin{lstlisting}
// ----- Develop from Scratch -----
// alpha -> exponential prefactor with the dimension of L^{-2}
// freelength -> rest length of binding crosslink
// D -> diameter of rod crosslink is binding to 
// construct positive lookup Baobzi Family object
Chebcoll bbcoll(0, alpha, freelength, D, 1);
// approximate functions
bbcoll.createBaobziFamily(1); 
// globally search domain to find range of integral in each grid
auto scanLoader = bbcoll.scanGlobalDomain(0, 1);
std::vector<std::vector<double>> tempkk = scanLoader.first;
std::vector<double> integralMinMax = bbcoll.intMinMax(tempkk); 
// construct reverse lookup Baobzi Family object
// use default output path -- can handle user input
Chebcoll bbcoll2(1, alpha, freelength, D, 3, 1e-1, 1e-3, integralMinMax);
// approximate functions
// unpack parameters
bbcoll2.createBaobziFamily(1, tempkk); 
// global search
bbcoll2.scanGlobalDomain(1, 1, testbound, 1, 0);
// using [.evalSinglePt(inval,0)] for single point evaluation; inval 2D coordinates

\end{lstlisting}

\begin{lstlisting}
// ----- Reconstruction -----
// reconstrct Baobzi Family object based on files in given path (currently using default path)
Chebcoll bbcoll2(3);
// global search
bbcoll2.scanGlobalDomain(1, 1, testbound, 1, 0);

\end{lstlisting}

\subsection{Normal Lookup}

\begin{lstlisting}
// ----- Develop from Scratch -----
Chebcoll bbcoll(1, alpha, freelength, D, 1);
bbcoll.createBaobziFamily(1); 
bbcoll.scanGlobalDomain(1,1);
    
\end{lstlisting}

\begin{lstlisting}
// ----- Reconstruction -----
Chebcoll bbcoll2(1);
bbcoll2.scanGlobalDomain(1, 1);

\end{lstlisting}

\bibliographystyle{splncs04}
\bibliography{mybib}

\end{document}